\documentclass[english,aps,prstper,preprint,showpacs]{revtex4-1}
\usepackage[T1]{fontenc}
\usepackage[latin9]{inputenc}
\usepackage{geometry}
\usepackage{amssymb}
\usepackage{graphicx}

\makeatletter

\providecommand{\tabularnewline}{\\}

\@ifundefined{textcolor}{}
{%
 \definecolor{BLACK}{gray}{0}
 \definecolor{WHITE}{gray}{1}
 \definecolor{RED}{rgb}{1,0,0}
 \definecolor{GREEN}{rgb}{0,1,0}
 \definecolor{BLUE}{rgb}{0,0,1}
 \definecolor{CYAN}{cmyk}{1,0,0,0}
 \definecolor{MAGENTA}{cmyk}{0,1,0,0}
 \definecolor{YELLOW}{cmyk}{0,0,1,0}
}

\makeatother

\usepackage{babel}

\begin{document}

\title{An Equity Investigation of Attitudinal Shifts in Introductory Physics}

\author{Adrienne Traxler}

\affiliation{Wright State University, 3640 Colonel Glenn Hwy., Dayton, OH 45435, USA}
\email{adrienne.traxler@wright.edu}

\author{Eric Brewe}

\affiliation{Florida International University, 11200 SW 8th St, Miami, FL 33199, USA}

\date{September 19, 2014}
\begin{abstract}
We report on seven years of attitudinal data using the Colorado Learning
Attitudes about Science Survey from University Modeling Instruction
(UMI) sections of introductory physics at Florida International University.
This work expands upon previous studies that reported consistently
positive attitude shifts in UMI courses; here, we disaggregate the data by
gender and ethnicity to look for any disparities in the pattern of
favorable shifts. We find that women and students from statistically
underrepresented ethnic groups are equally supported on this attitudinal
measure, and that this result holds even when interaction effects
of gender and ethnicity are included. We conclude with suggestions
for future work in UMI courses and for attitudinal equity investigations
generally.
\end{abstract}

\maketitle

\section{Introduction}

The University Modeling Instruction curriculum (UMI; \onlinecite{brewe_modeling_2008})
developed and studied at Florida International University (FIU) has
produced an uncommon pattern of consistently positive shifts in student
attitudes toward physics \citep{brewe_extending_2013}. The case for
studying student attitudes and epistemologies has been made at greater
length elsewhere \citep{brewe_extending_2013,redish_student_1998,national_research_counci_discipline-based_2012};
here, we will summarize those arguments, but largely take as a given
that improving students' attitudes toward physics is one relevant
dimension of success for a curriculum. However, education researchers
must be cautious of overgeneralizing results, and one such overreach
is to claim that a benefit is received by all students when in fact
it only accrues to those from majority groups. FIU, a Hispanic-serving
institution with a large fraction of women in the calculus-based Modeling
sections, provides an important opportunity to investigate this aspect
of the UMI curriculum with a diverse student body. Section \ref{sec:Background}
discusses the context of gap-based analyses in education research,
outlining some of the pitfalls of this approach and why we have chosen
it here, and also summarizes some of the most relevant results on
attitude surveys. Section \ref{sec:Methods} outlines the context
of data collection and the research questions considered. Section
\ref{sec:Results} summarizes our results, and Section \ref{sec:Discussion-and-Conclusions}
concludes with suggestions for future equity investigations of attitudinal
or conceptual measures, cautioning to avoid forms of ``gap-gazing''
that can further marginalize underrepresented groups.

\section{Background\label{sec:Background}}

\subsection{Gaps analyses}

Examination of performance differences, or looking for ``gaps''
between groups, is not without controversy in education research.
As outlined by \citet{gutierrez_gap-gazing_2008} in mathematics and
\citet{danielsson_gender_2010} in physics, gaps analyses run the
risk of essentializing student identities by overgeneralizing (e.g.,
``all women...''). Guti\'errez argues that gap analyses often implicitly
reinforce a deficit model in which students' differences are presumed
to be the result of inadequacies in preparation, skill, or ability.
Further, she argues, this frames students from different backgrounds
in opposition with one another. \citet{lubienski_gap_2008}, on the
other hand, contends that investigations of gaps are critically important
to inform education policy and that it would be ``irresponsible''
to stop making gaps analyses. Following on Lubienski, we feel that
it is not just valuable but essential for teachers and curriculum
developers to question whether the benefits of instruction are distributed
equitably among statistically underrepresented and majority student
groups.

Some gap-based analyses, when thoughtfully conducted, have deepened
our understanding of the mechanisms behind systemic performance differences
on traditional academic measures. One key example is stereotype threat,
originally uncovered when testing different framings of a difficult
verbal test given to white and African American students \citep{steele_stereotype_1995}.
This landmark study and many following (for one review, see \citealp{aronson_stereotypes_2005})
reveal a previously invisible barrier for women and students from
statistically underrepresented racial and ethnic groups. Aware of
negative stereotypes about their groups and invested in disproving
them, these students face extra cognitive load from their awareness,
and often show performance drops in the very subjects where they care
the most \citep{aronson_when_1999}. 

Stereotype threat research has led to a richer understanding of how
to frame classroom tasks in a manner that better supports all students.
This work, including some in physics education research \citep{miyake_reducing_2010},
would not have been possible without a willingness to investigate
the causes of systematically observed performance differences between
groups. Indeed, while Guti\'errez outlines pitfalls of gaps analyses,
she also gives suggestions for avoiding them \citep{gutierrez_gap-gazing_2008}.
These suggestions include include a greater focus on intervention
work and on effective teaching and learning environments. In the spirit
of the latter category, we focus our attention on data collected from
University Modeling Instruction classes at FIU. 

\citet{rodriguez_impact_2012} discuss three predominant models of
equity in the context of physics education research: Equity of Fairness,
Equity of Parity, and Equity of Individuality.  Under the Equity of
Fairness model, students from all populations should experience similar
gains or losses. Equity of Fairness models would preserve pre-existing gaps. In the Equity of Parity model, students from one population might enter with lower scores on some measure, but all should leave with the same score distribution. Interventions striving for gap closing work from an Equity of Parity model. Finally, Equity of
Individuality investigations explicitly avoid group comparisons and
instead focus on understanding individual excellence. Gap-based analyses
are unable to speak to this model, but may still provide important
insight to equity of fairness or equity of parity questions. Research
such as this paper, which explores differences in attitudinal shifts
between groups, is relevant to Equity of Parity and Equity of Fariness models.

Previous work has outlined the epistemological goals of the UMI curriculum,
which frames modeling as the key activity of scientists \citep{brewe_modeling_2008,hestenes_toward_1987}.
UMI classes have shown favorable student outcomes in conceptual understanding
\citep{brewe_toward_2010}, self-efficacy \citep{sawtelle_positive_2010,sawtelle_creating_2012},
in student social network measures \citep{brewe_changing_2010}, and
in student attitudes towards physics \citep{brewe_extending_2013}
and engaging in physics \citep{goertzen_expanded_2013}. We expand on
the latter work here by examining whether these attitudinal gains
are shared equally by women and by students from black, Hispanic,
Native American, and Pacific Islander ethnicities. As of this paper's
writing, all four ethnic groups are statistically underrepresented
in the sciences and in physics, relative to the demographics of the
United States population%
\footnote{Data from the American Physical Society (http://www.aps.org/programs/education/statistics/),
sourced from the IPEDS Completion Survey.%
}. In the text, we will adopt this language of ``statistically underrepresented,''
to avoid the deprecating connotations of ``underrepresented minorities''
and also to more accurately reflect FIU's status as a predominantly
Hispanic institution.

\subsection{Student attitudes}

A variety of studies now document student attitudes in introductory
university physics \citep{halloun_views_1997,redish_student_1998,adams_new_2006},
and the effects of students' attitudes and epistemologies on their
conceptual gains \citep{halloun_views_1997,perkins_correlating_2005},
use of content knowledge \citep{lising_impact_2005}, and choice of
courses and majors \citep{perkins_correlating_2005,perkins_who_2010}.
However, these results are not always reported through a lens of demographic
factors. While attention has been paid to gender differences on conceptual
inventories (see \citealt{madsen_gender_2013} for a recent review),
comparatively little attention has been paid to differences between
majority and statistically underrepresented ethnic groups. Studies
of student attitudes are less common than research on conceptual gains,
and most have not disaggregated by demographic factors. However, some
reported CLASS data has shown more favorable pretest attitudes and
shifts for men \citep{adams_new_2006,kost_characterizing_2009}. 

Despite a dearth of research on differential attitudes toward physics
or science generally, the research on stereotype threat introduced
above cautions us that attitudinal differences are very salient for
students from statistically underrepresented groups. A serious long-term
consequence of stereotype threat is the filtering effect it applies
to participation: students from negatively-stereotyped groups, over
time, often disidentify with the threatened area \citep{steele_threat_1997}.
As a result, differentially lower initial attitudes or negative attitudinal
shifts may be an important warning to instructors of disengagement
in students from threatened groups. 

Research from the University of Colorado has shown that initial (pre-university
instruction) student attitudes are strongly correlated with pursuing
a physics major \citep{perkins_who_2010}. It remains an open question
whether positive shifts in attitude show a similar longitudinal signal
(open, in part, because demonstrating consistently positive shifts
has itself been a substantial barrier). However, at a more fine-grained
scale, a positive shift in attitudes toward physics learning has been
linked with more central membership in the physics community in FIU's
rapidly growing physics major population \citep{goertzen_expanded_2013}.

We have ample motivation to examine
patterns of positive attitudinal shifts, as potential signals of growing
student investment and participation in physics. However, to accurately
report promising findings, we must also ask whether any such benefits
are equally received by all groups of interest. In this paper, we
investigate pre-course to post-course attitude scores and shifts for
students in calculus-based Physics I (mechanics) courses. From previous
work, we know that the University Modeling Instruction courses are
equitable by the Equity of Fairness model for Force Concept Inventory
gains by ethnicity, but not by gender \citep{brewe_toward_2010}.
Here we extend the equity question to attitudinal shifts. This investigation
contributes to the knowledge base on impacts of student attitudes
by first exploring differential attitudes across statistically underrepresented
student groups, and then by asking how instruction shifts student
attitudes among these groups.

\section{Methods\label{sec:Methods}}

FIU is a large, minority-serving institution (54,000 students, 61\%
Hispanic, 13\% black, in Spring 2014) with a primarily commuter student
body. Over the past ten years, the Physics Education Research Group
has guided a series of structural reforms in the introductory physics
courses, including the addition of University Modeling Instruction
sections of the calculus-based sequence.  

The data presented in this paper is drawn from introductory physics
I courses and was collected from the Fall 2007 to Fall 2013 semesters.
The Colorado Learning Attitudes about Science Survey\citep{adams_new_2006}
was administered on paper at the beginning and end of each term and
filtered for matched student responses. Table \ref{tab:Demographics}
shows the demographics of the student sample. The gender ratio is
much closer to parity in UMI sections than in traditional lecture
physics courses at FIU, while the distribution of students' ethnic
representation is very similar between the two course formats.

\begin{table}
\begin{centering}
\begin{tabular}{|c|c||c|c|c|c|c|c|}
\hline 
\multicolumn{2}{|c||}{Gender} & \multicolumn{2}{c|}{Ethnicity (SR)} & \multicolumn{4}{c|}{Ethnicity (SUR)}\tabularnewline
\hline 
F & M & ASIAN & WHITE & AMIND & BLACK & HISPA & PACIF\tabularnewline
\hline 
\hline 
51.5\% & 48.5\% & 6.1\% & 8.0\% & 0.4\% & 8.3\% & 76.9\% & 0.4\%\tabularnewline
\hline 
\end{tabular}
\par\end{centering}

\protect\caption{Demographics of University Modeling Instruction sections in the sample
(N=264). Percentages are given for gender and for ethnic representation,
grouping by statistically well- or over-represented (SR) and statistically
underrepresented (SUR).\label{tab:Demographics}}
\end{table}

We look for pretest, post-test, and shift differences between students
who are statistically well- or over-represented in physics (male,
Asian, and white students) and those who belong to statistically underrepresented
groups (female, black, Hispanic, Native American, and Pacific Islander
students). We seek to answer two research questions: 
\begin{enumerate}
\item Do gender or ethnic representation influence students\textquoteright{}
percentage of expert-like CLASS responses in University Modeling Instruction? 
\item Is there an interaction between gender and ethnic representation?
\end{enumerate}
To address the first question, we disaggregate student pretests, post-tests,
and shifts in percentage favorable responses on the CLASS. In addition
to checking for statistically significant differences in these values
between groups, we follow \citet{rodriguez_impact_2012} in looking
for significant effect sizes. Measured using Cohen's $d$ \citep{cohen_power_1992}:

\begin{equation}
d=\frac{\mu_{2}-\mu_{1}}{\sigma_{pooled}}
\end{equation}
the effect size provides an indicator of ``practical significance,''
and thus serves as a necessary accompaniment to statistical significance
when reporting claims about gaps between groups \citep{rodriguez_impact_2012}.

The second question occurs because the intersection of gender and
racial or ethnic identity is known to pose additional challenges for
women of color in the sciences \citep{ong_inside_2011}. To address
this point, we use a linear regression model including an interaction
term for gender and ethnicity, and investigate whether it explains
a significant amount of the variance in post-course attitudes.

\section{Results\label{sec:Results}}

\begin{figure}
\includegraphics[width=0.9\columnwidth]{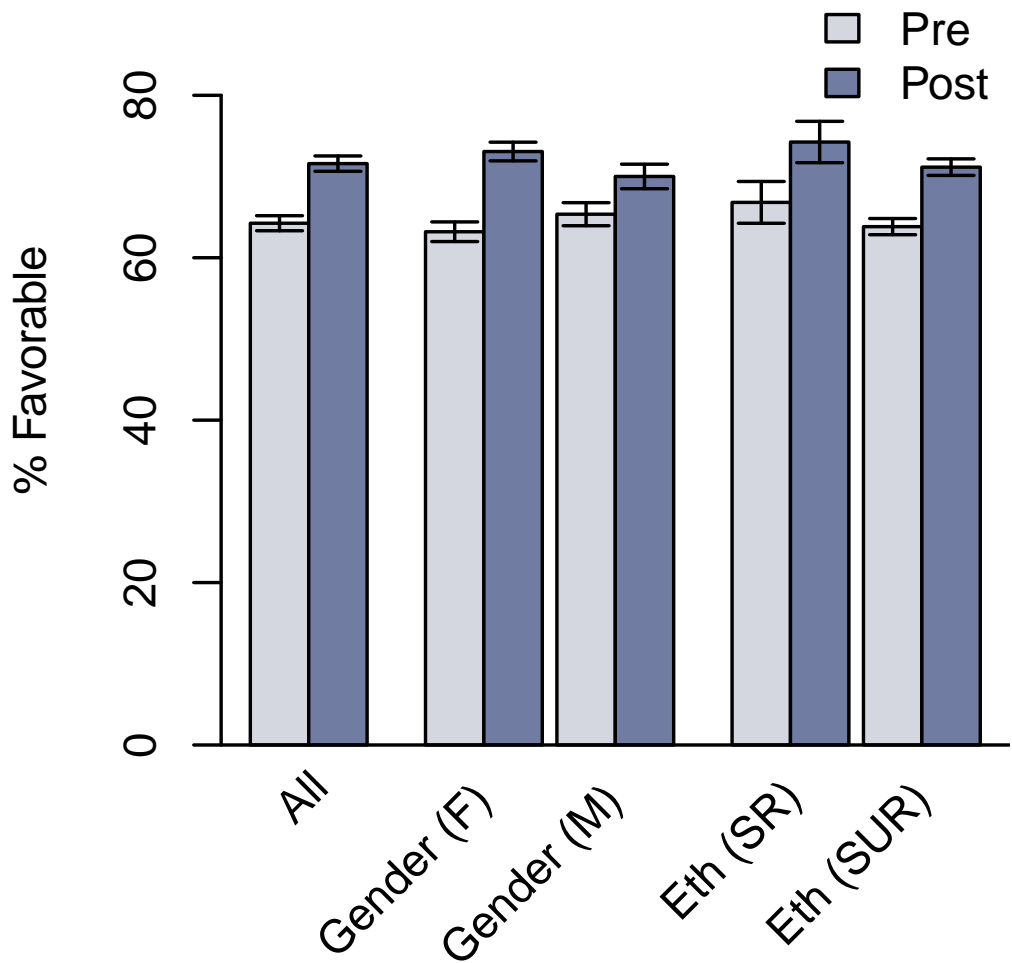}

\protect\caption{(Color online) Overall favorable average CLASS scores for Modeling Instruction sections.
Bars show standard error of the mean.\label{fig:PrePost}}
\end{figure}

\begin{figure}
\includegraphics[width=0.9\columnwidth]{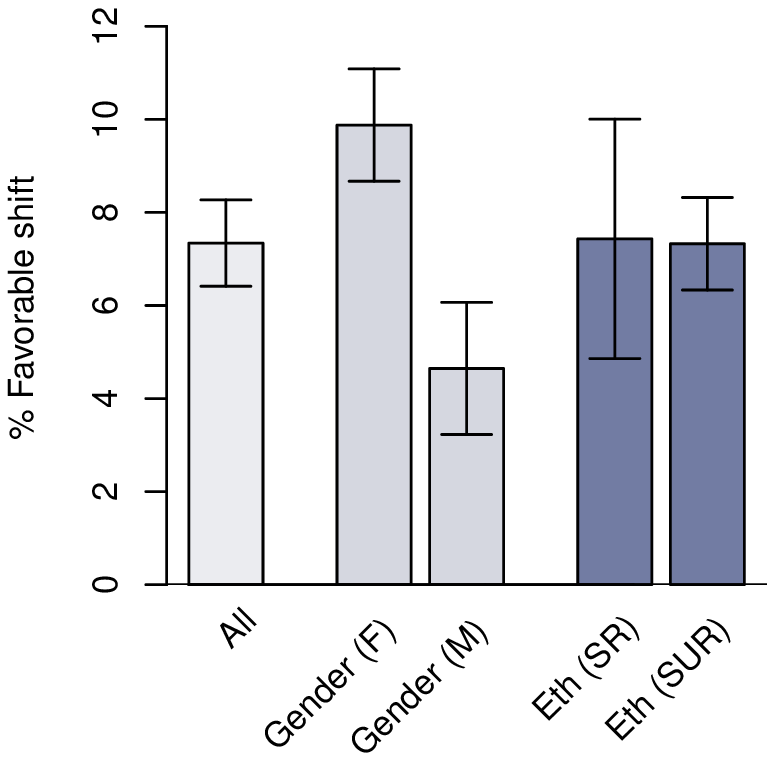}

\protect\caption{(Color online) Average shift in overall favorable percentage CLASS score, with bars
showing standard error. Shifts are shown overall and then disaggregated
by gender (male or female) and ethnic representation (statistically
represented or statistically underrepresented).\label{fig:Shifts}}
\end{figure}

Figure \ref{fig:PrePost} shows the significant and positive differences
between pre- and post-course responses in the Modeling classes. Disaggregating
by gender and by ethnic representation, we see that all subgroups
show significant positive shifts. Figure \ref{fig:Shifts} elaborates
on the disaggregated results by showing percentage favorable shifts
for all students, by gender, and by ethnic representation. 

\begin{figure}
\includegraphics[width=0.9\columnwidth]{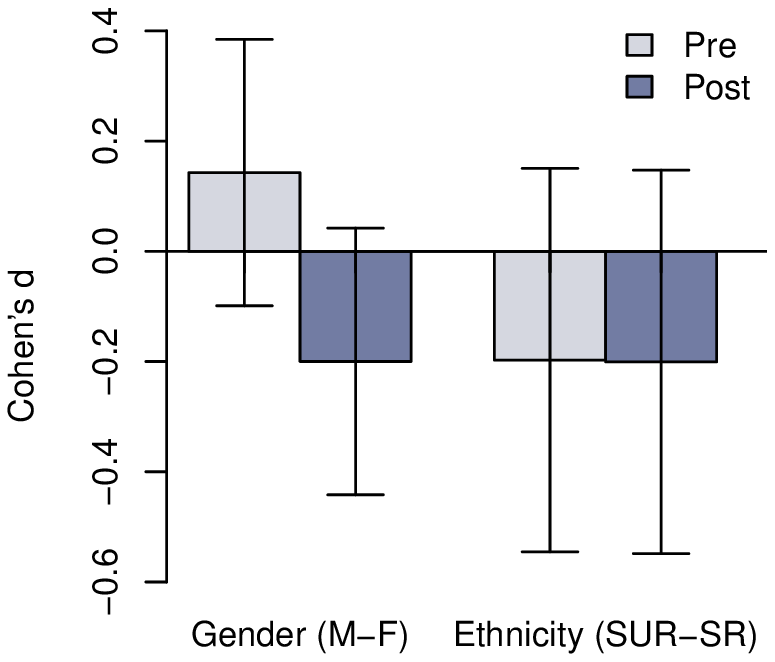}

\protect\caption{(Color online) Effect sizes of disaggregated groups. In all cases, differences between
group means are at or below the threshold for small effect size, with error
bars crossing the axis indicating no meaningful effect. \label{fig:Effect-sizes}}
\end{figure}

Figure \ref{fig:Effect-sizes} shows the effect sizes, Cohen's $d$,
of group differences on pre- and post-test. We see that for both gender
and ethnicity, on pre- and post-course administrations of the CLASS,
the effect sizes of the differences are small ($d\lesssim0.2$) and
the error bars overlap zero. This overlap indicates that there is
no meaningful difference between the pre- and post-course means for
men compared to women, or statistically represented compared to statistically
underrepresented ethnicities. 

Finally, to check for possible interactions of gender and ethnicity
that might be overlooked when considering each factor individually,
we use a linear regression model:

\begin{equation}
\text{Post} \sim \text{Pre}+ \text{Gender}+\text{EthRep}+\text{Gender}\times \text{EthRep}
\end{equation}

Here, Post and Pre represent the overall percent favorable scores,
Gender is coded as F or M%
\footnote{We acknowledge that this gender binary is a simplification, but our
demographic data comes from the university database, which presently
only allows these two choices.%
}, and EthRep is coded SR or SUR for statistically represented or underrepresented
ethnic groups respectively. 

Fitting this model to the sample of 264 students, we find that only
the coefficient for Pre is significant: $\beta_{\text{Pre}}=0.57$,
95\% CI = (0.46, 0.67), $p<0.01$. For the full model, $R^{2}=0.32$,
indicating that substantial variance remains unexplained. Neither
gender nor ethnic representation, or the interaction between them,
were significant predictors of post-course expert-like beliefs once
a student's pre-course beliefs were accounted for. This result confirms
the non-significant effect sizes found above, and clarifies that there
is no detectable gender-ethnicity interaction that was hidden by splitting
the data along those categories.

\section{Discussion and Conclusions\label{sec:Discussion-and-Conclusions}}

Previous studies of student conceptual gains in introductory physics
have pointed to a disparity between male and female students \citep{brewe_toward_2010,madsen_gender_2013}.
Results vary on whether these gaps persist in reform-based classes,
where various features of the learning environment might be expected
to support traditionally marginalized students. Although there is
important debate about the degree to which ``gap-gazing''
is useful or appropriate in education research, a gender or ethnicity-divided
difference in gains is troubling because it suggests that not all
students are receiving the claimed benefits of reform efforts. 

In the attitudinal study reported here, the picture is somewhat different
than for conceptual measures. Returning to our research questions: 

\emph{1. Do gender or ethnic representation influence students\textquoteright{}
percentage of expert-like CLASS responses in University Modeling Instruction?}
There is no evidence that either female students, or those from statistically
underrepresented ethnicities, have either lower or higher pre-course,
post-course, or shifts in percentage of favorable beliefs on the CLASS.
Closer examination of the score distributions does show some evidence
of a ceiling effect on the post-course CLASS, but it does not appear
that the effect is stronger for the traditionally-majority groups
(which, if the case, might artificially suppress a gap). It would
be very useful to disaggregate the scores by gender and ethnicity
for a broader sample of classes, where high pretest scores are less
prevalent, and for non-Modeling courses (more on this below).

\emph{2. Is there an interaction between gender and ethnic representation?}
In a linear model of post-course attitude scores where gender, ethnicity,
and their interaction are included, none of these coefficients are
statistically significant. Only students' pre-course attitudes are
a significant predictor in the model, and even this term only accounts
for 32\% of the total variance in post-course attitude scores. So
far as we can detect with this data, women from statistically underrepresented
ethnicities have a similar pre- and post-course attitude profile as
their peers in other groups.

Revisiting the two models of equity discussed in Section \ref{sec:Background},
the Modeling classes are supportive of student attitudes in the equity
of fairness sense, where all groups show similar gains. As no pre-course
differences in distribution existed, nor did traditional majority
groups show disproportionate gains, Modeling is also supportive of
student attitudes by the Equity of Parity model.

As noted above, one possible explanation for FCI gender gaps is stereotype
threat, which is known to depress the performance of women and students
from underrepresented ethnic groups on many academic tasks. However,
a key component of the threat is perceived risk of doing badly on
a task where one will be judged. An attitude survey, where students
are asked to rate their beliefs rather than to choose one correct
answer, may be perceived as a less failure-prone task and thus not
trigger the threat. However, this explanation does not account for
the lack in our sample of a gender difference in attitudes, which
has been observed in other CLASS studies. Additionally, the pre-course
attitudes for students entering the Modeling classes are already very
favorable, above lecture students at the same institution and at the
high end of typical pre-course scores reported for the CLASS (\citealt{adams_new_2006},
Table VIII).

Possible explanations include greater student buy-in at the beginning
of the semester, as students must apply and be selected by lottery
due to the popularity of the UMI sections. A related hypothesis is
that the same informal network of peers that passes information about
the course may also confer a higher expectation of success, leading
to a self-efficacy boost that registers on the CLASS. To help account
for the first possibility, a more comprehensive attitude survey of
lecture students in the same cohort would be useful: tracking students
who unsuccessfully applied to Modeling sections, and comparing their
CLASS pre-course scores with those who found seats in Modeling, could
detect whether the UMI classes somehow attract more ``physics people.''
Observationally however, this is somewhat unlikely, as many UMI students
are on pre-medical paths and have no initial interest in a physics career. 

Returning to the question of ``gap-gazing,'' looking for performance
differences between groups should be done carefully, because it risks
problematizing already marginalized students. But until the field
of physics accurately reflects the diverse talents of the population,
and until effects such as stereotype threat are no longer detectable,
it is important for education researchers to address whether their
reforms truly are for all students. Building on this awareness, a
constructive way to address the problems of underrepresentation in
science is to examine successful curricula and learning environments
so that lessons may be drawn from positive examples. 

In this work, we have examined the favorable attitudinal shifts reported
in UMI courses, asking whether they are equitable among students of
different genders and ethnicities. We find that they are, and somewhat
surprisingly, that this is true even on a pre-course attitude survey
where more negative attitudes have been reported for women in other
studies. While it would be unreasonable to attribute this pre-course
parity to the UMI curriculum, it suggests that a fruitful dimension
for research to expand is beyond the boundaries of classroom pre-
and post-tests, to investigate the learning networks and communities
that may transmit information and expectations to future students.
The results reported here, taken together with previous FCI and odds
of success comparisons for the same courses \citep{brewe_toward_2010},
also caution against taking any one test score---attitudinal, conceptual,
or otherwise---as the solitary measure of a student group. Multiple
measures of success are needed to understand, measure, and value the
many things that students learn in physics courses.

\begin{acknowledgments}
This manuscript is based upon work supported by the Howard Hughes
Medical Institute'
s Science Education Grant No.\ 52006924
and by the National Science Foundation under Grant No.\ 0802184 and DUE 1140706. The
authors would like to thank the students who participated in this
study and the FIU Physics Education Research Group for their insight.
\end{acknowledgments}

\bibliographystyle{apsrev}

\end{document}